\documentclass[aps,prd,twocolumn,superscriptaddress,showpacs]{revtex4}
\usepackage{natbib,hyperref,ifthen} 
\usepackage{graphicx}
\usepackage{dcolumn}
\usepackage{bm}
\usepackage{natbib}
\usepackage{multirow}
\usepackage{epsfig}
\usepackage{amsmath}

\newcommand{\beq}{\begin{equation}}
\newcommand{\eeq}{\end{equation}}
\newcommand{\beqa}{\begin{eqnarray}}
\newcommand{\eeqa}{\end{eqnarray}}

\newcommand{\be}{\begin{equation}}
\newcommand{\ee}{\end{equation}}

\begin{document}

\title{A New Figure of Merit for Dark Energy Studies}

\author{An\v{z}e Slosar} 
\affiliation{Brookhaven National Lab, Bldg 510A, Upton NY 11973}

\date{April 1$^{\rm st}$, 2010}

\begin{abstract}
  We introduce a new figure of merit for comparison of proposed dark
  energy experiments. The new figure of merit is objective and has
  several distinct advantages over the Dark Energy Task Force Figure
  of Merit, which we discuss in the text.
\end{abstract}


\maketitle

\setcounter{footnote}{0}

\section{Introduction}
\label{sec:introduction}

The measurements of the Cosmic Microwave Background (CMB)
anisotropies, most notably by the Wilkinson Microwave Anisotropy Probe
(WMAP) mission \cite{Jarosik:2010iu}, but also by ground based and
balloon borne experiments, such as VSA \cite{2003MNRAS.341L..29S},
CbI \cite{CBI:04}, AcbaR \cite{ACBAR:07}, SPt \cite{Reichardt:2009ys},
QuaD \cite{Brown:2009uy} are in spectacular agreement \cite{Melch} by
predictions of the standard cosmological model. Measurements of the
low-redshift universe including data from large spectroscopic surveys like
SDSS \cite{abazajian:09,Schleg} analysed using a variety of methods
\cite{Percival:2009xn,2006PhRvD..74l3507T}, measurements of the
luminosity distance to type Ia supernovae
\cite{riess:09,2008ApJ...686..749K} and the galaxy-galaxy lensing
\cite{2008MNRAS.390.1157R,2006ApJ...640..691V} strengthen the standard
picture. These datasets provide constraints on the cosmological model
using a variety of techniques with very different systematic issues,
which nevertheless converge, within the error-bars, on the standard
model of cosmology.

Perhaps the most surprising aspect of the standard cosmological model
is the overwhelming evidence that the Universe is undergoing
accelerated expansion.  This expansion can be most easily explained in
terms of cosmological constant $\Lambda$. In fact, from the
theoretical perspective, the cosmological constant is perhaps a very
natural phenomenon, whose small size only illustrates our poor
understanding of the fundamental theory of the Universe \cite{Bianchi:2010uw,Bousso:2009ks}.

However, motivated by the need to establish new and much needed gaps
in literature, various authors have considered alternatives to
cosmological constant, which most often include new degrees of
freedom. Such models generically predict effective equation of state
$w=p/\rho$ for these novel components of the Universe that can deviate
from the value for the cosmological constant, namely $w=-1$ by
arbitrarily small amounts \cite{Caldwell:1997ii}. Concurrently with
these efforts, an industry of phenomenological models has been
established. One of the most popular descriptions for the dynamical
dark energy is the $w_0$-$w_a$ parametrisation, in which the equation
of state is postulated to evolve with cosmological scale factor
$a=(1+z)^{-1}$ (where $z$ is redshift) as \cite{2003PhRvL..90i1301L}
\begin{equation}
  w(a) = w_0 + (1-a) w_a.
\end{equation}

At the same time, motivated by the need to attract funding,
experimentalists have begun proposing various experiments that will
measure the value of $w$ and its derivatives with an ever increasing
precision. In fact, designing cosmological experiments around
measuring neutrino masses from cosmology has a distinct disadvantage
in that sum of neutrino mass eigenstates has a lower limit given by
the ground-based experiments
\cite{kamland:2008ee,2004PhRvL..92r1301A,2004PhRvL..93j1801A}. The
same holds true for constraining theories of inflation by measuring
the running of the spectral index, which is expected to be of
$O(10^{-4})$ in the simplest inflationary models, given the current
limits on the tilt of the spectral index $n_s\sim0.96$
\cite{Komatsu:2010fb}.  Measuring $w$ poses no such difficulties:
since there is a strong theoretical prejudice that $w=-1$, one can
hope to improve limits on deviation from the cosmological constant
value for decades to come.

An interesting question, worth every penny of scientific funding, is
the question of comparison of various dark energy experiments. Several
methods have been established for this purpose, the most common is the
Dark Energy Task Force (DETF) Figure of Merit (FoM)
\cite{2006astro.ph..9591A}. In this paper we propose a new metric,
that has several advantages. We discuss the DETF FoM and our new metric
in Section \ref{sec:new-figure-merit}. We conclude in Section
\ref{sec:Conclusions}. Finally, we also note that clusters of galaxies
are the most massive gravitationally bound objects in the Universe.

\section{A new figure of merit: $\mathbf{\ddot{\phi}}$}
\label{sec:new-figure-merit}

\begin{figure}
  \centering
  \includegraphics[width=\linewidth]{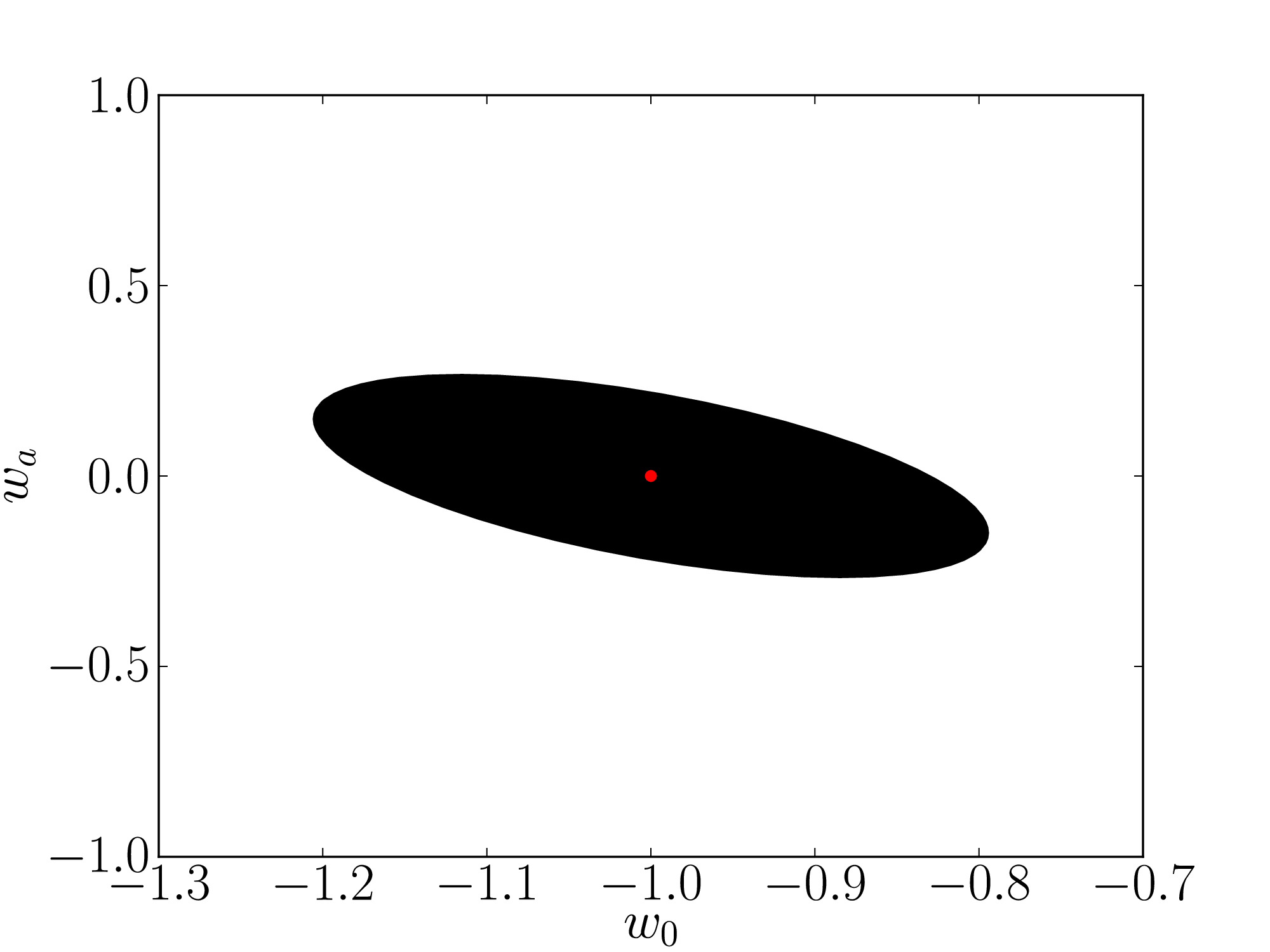}
  \includegraphics[width=\linewidth]{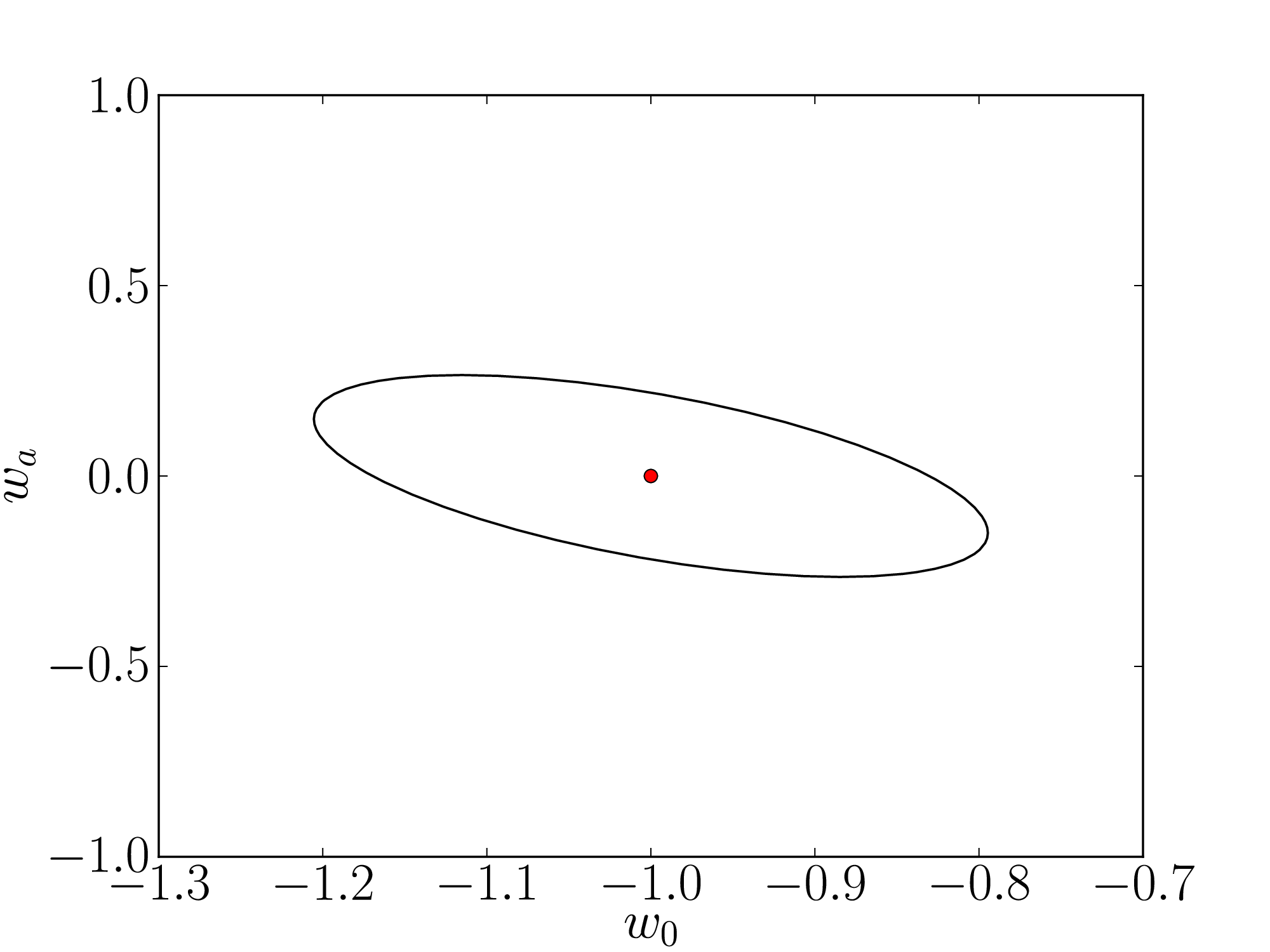}
  \caption{This figure illustrates the DETF FoM (top panel) and
    $\mathbf{\ddot{\phi}}$ (bottom panel). The DETF FoM is inversely
    proportional to the amount of ink necessary to print the ellipse
    in the top panel, while $\mathbf{\ddot{\phi}}$ is inversely proportional to
  the amount of ink necessary to plot the ellipse int the bottom
  panel (in the limit of infinitely thin lines).  Red dot denotes the
  fiducial model used in this work.\label{fig:1}}
\end{figure}

We start by considering the well-established DETF FoM. This figure of
merit is defined as 
\begin{equation}
  {\rm DETF\ FoM} = \left( {\rm det} C \right)^{-1/2},
\end{equation}
where $C$ is the $2\times2$ covariance matrix of the errors on the
$w_0$-$w_a$ plane
\begin{equation}
C = \left( \begin{array}{cc}
\sigma^2_{w_0w_0} & \sigma^2_{w_0w_a} \\
\sigma^2_{w_0w_a} & \sigma^2_{w_aw_a}  \\
\end{array} \right).
\end{equation}
This figure of merit can be interpreted as the inverse of the area of
the error ellipse on the $w_0$-$w_a$ plane. Now come our ingenious and
novel idea. We introduce our new figure of merit, whose value is
proportional to the inverse of the \emph{circumference} of the error
ellipse on the $w_0$-$w_a$ plane. To calculate this quantity, we first
note that the semi major and semi minor axes of the error ellipse are
given by the square root of the eigen-vectors of the covariance
matrix:
\begin{equation}
  r^2_{\pm} = \frac{1}{2}\left(\sigma^2_{w_0w_0}+ \sigma^2_{w_aw_a} \pm \sqrt{\left(\sigma^2_{w_0w_0}-\sigma^2_{w_aw_a}\right)^2+4\sigma_{w_0w_a}^2}\right).
\end{equation}
The circumference is then given by 
\begin{equation}
  s = 4 r_+ E(e), 
\end{equation}
where  eccentricity of the ellipse is given by
\begin{equation}
  e = \sqrt{1-\frac{r_-^2}{r_+^2}},
\end{equation}
and $E(e)$ is the complete elliptic integral of the second kind.  Our
new figure of merit is then given by
\begin{equation}
  \mathbf{\ddot{\phi}} = s^{-1}.
\end{equation}
The symbol for our new figure of merit is $\mathbf{\ddot{\phi}}$ which is to be
pronounced as \textbf{ph\"u} and not ``phi double-dot''. We illustrate the two
figures of merit in the Figure \ref{fig:1}.

\begin{table}
  \centering
  \begin{tabular}{ccc}
    Experiment &   ${\rm DETF\ FoM}$ & $\mathbf{\ddot{\phi}}$ \\
    \hline
    Somewhat good experiment & 95 & 39 \\
    Very good experiment & 403 & 80 \\
    CNDEM\footnote{Lorentz violating Chuck Norris in space, breathing
      aether and watching galaxies with his naked
      eyes. For Chuck Norris using specs, add 40\% to the DETF FoM and 18\%
      to $\mathbf{\ddot{\phi}}$}. & 845 & 132 \\
    FMIE\footnote{Fisher Matrix Itself Experiment} & 1693 & 180\\
  \end{tabular}
\caption{Comparison of standard and improved Figures of Merit for a
  selection of proposed experiments. Systematics issues were ignored
  when calculating these Figures of Merit, due to difficulties in
  modelling them. This table illustrates superiority of $\mathbf{\ddot{\phi}}$
  over DETF FoM. \label{tabl:1}}
\end{table}

Our new and improved figure of merit has several advantages over DETF FOM:
\begin{itemize}

\item When two experiments have the same DETF FoM, the $\mathbf{\ddot{\phi}}$
  quantity will favour one with less correlated errors on the
  $w_0$-$w_a$ plane.  Since this plane is well motivated by the
  fundamental physics, it is clear that uncorrelated errors should be
  favoured;

\item Area grows proportionally to the square of the linear
  dimensions, while circumference grows only linearly. This makes
  $\mathbf{\ddot{\phi}}$ more linear;

\item Calculation of the new figure of merit entails calculating
  elliptic integrals of the second kind, which makes the method more
  scientific;

\item Correct pronunciation of $\mathbf{\ddot{\phi}}$ allows one to shower the
  opponents face in one's saliva, thus quickly and effectively
  dispersing any doubts about the superiority of the experiment
  proposed by the speaker.
\end{itemize}

We also note that both figures of merit can be generalised to models
with more than two parameters. For the DETF FoM, this has been
performed in \cite{Wang:2008zh}, but our new figure of merit is
considerably more complicated and so we defer this work for future
publication.

We compare the two figures of merit in Table \ref{tabl:1} for a couple
of proposed experiments. We note that the experiments which are
further into the future have better figures of merit. We also note
that the more expensive experiments have better figures of merit. The
table demonstrates the superiority of the new figure of merit.

\section{Conclusions and Discussion}
\label{sec:Conclusions}

In this paper we have introduced a new figure of merit, $\mathbf{\ddot{\phi}}$,
which is proportional to the inverse of the circumference of the error
ellipse.

As discussed in the text and we discuss it here again, the new figure
of merit has several advantages over the old one. You and your dog
should use it. If you do not use it and think it is a pointless
number, you should nevertheless cite this paper or I will write you
hassling emails.  If worse come to worse, I'll resort to the crowbar
and smash your 30 inch liberal screen.

This work opens clear avenues for further research. The quantity
$\mathbf{\ddot{\phi}}$ can and should be calculated for many future experiments
and further compared to the DETF FoM. Rigorous extension of this work
into models of dark energy with more than two parameters remains to be
performed. Theoretical basis for similarities and differences between
the two figures of merit should be established and elaborated.
Different parametrisation of the dark-energy should be integrated into
the new figure of merit resulting in a multitude of useful figures of
merit. The best in the science of figures of merit has yet to come!

\section*{Acknowledgements}

AS acknowledges useful discussions with the usual suspects. He really
wants a cat to discuss dark energy and play checkers. This work was in
no way whatsoever supported in part by the U.S. Department of Energy
under Contract No. DE-AC02-98CH10886.


\bibliographystyle{arxiv}
\bibliography{cosmo,cosmo_preprints,fR}
\end{document}